\documentclass[%
preprint,
 amsmath,amssymb,
 aps,
]{revtex4-1}

\usepackage{graphicx}
\usepackage{dcolumn}
\usepackage{bm}
\usepackage{multirow}

\usepackage{graphicx}
\usepackage{epstopdf, epsfig}
\usepackage[retainorgcmds]{IEEEtrantools}
\usepackage{booktabs}

\usepackage{lineno}
\usepackage{mathtools}
\usepackage{tabularx}
\usepackage{xcolor}

\begin{document}


\title{Sensitivity of vortex pairing and mixing to initial perturbations in stratified shear flows}

\author{Wenjing Dong}
 \altaffiliation{Department of Civil Engineering, University of British Columbia,
		Vancouver, BC V6T 1Z4, Canada}

\author{E. W. Tedford}
 \altaffiliation{Department of Civil Engineering, University of British Columbia,
		Vancouver, BC V6T 1Z4, Canada}

\author{M. Rahmani}
 \altaffiliation{Department of Mathematics, University of
		British Columbia, Vancouver, BC V6T 1Z2, Canada}
\author{G. A.  Lawrence}
 \altaffiliation{Department of Civil Engineering, University of British Columbia,
		Vancouver, BC V6T 1Z4, Canada}

\date{\today}

\begin{abstract}
The effects of different initial perturbations on the evolution of stratified shear flows  that are subject to Kelvin-Helmholtz instability and vortex pairing have been investigated through Direct Numerical Simulation (DNS).  The effects of purely random perturbations of the background flow are sensitive to the phase of the subharmonic component of the perturbation that has a wavelength double that of the Kelvin-Helmholtz instability.  If the phase relationship between the Kelvin-Helmholtz mode and its subharmonic mode is optimal, or close to it,  vortex pairing occurs.  Vortex paring is delayed when there is a phase difference, and this delay increases with increasing phase difference.  In three dimensional simulations vortex pairing is suppressed if the phase difference is sufficiently large, reducing the amount of mixing and mixing efficiency.  For a given phase difference close enough to the optimal phase, the response of the flow to eigenvalues perturbations is very similar to the response to random perturbations. In addition to traditional diagnostics, we show quantitatively that a non-modal Fourier component in a random perturbation quickly evolves to be modal  and describe the process of vortex pairing using Lagrangian trajectories.

\end{abstract}
\maketitle

\section{Introduction}
Fluids are often stably stratified in the atmosphere, ocean, and lakes, due to temperature or salinity or both. The existence of shear (vertical variations in the horizontal currents) may give rise to instabilities in these otherwise stably stratified flows. Kelvin-Helmholtz (KH) instabilities, also called Rayleigh instabilities in homogeneous fluids, are one of the most widely known shear instabilities. KH instabilities have been studied extensively in both homogeneous and stratified fluids using laboratory experiments \citep[e.g.][]{Thorpe73, Browand73, Winant74}, field observations \citep{Seim94, Moum03, Geyer10}, and numerical simulations \citep[e.g.][]{Patnaik76, Klaassen85, Caulfield00, Staquet00, Mashayek12a, Mashayek13, Rahmani14, Salehipour15}. They are characterized by  two-dimensional periodic elliptic vortices called KH billows, which are connected by thin tilted braids of high strain rate \citep{Corcos76}. 

KH instabilities  are susceptible to several secondary instabilities, e.g. vortex pairing \citep{Browand73, Winant74, Koop79, Ho82}, convective core instability due to the overturn of fluid caused by the roll-up \citep[e.g.][]{Klaassen85, Caulfield00}, and instabilities that are located in braid regions and extract energy from the mean shear or strain \citep[see][]{Mashayek12a}. Which secondary instabilities exist or dominate depends on non-dimensional parameters governing the flows, i.e. Reynolds number, Richardson number, and Prandtl number. \citet{Klaassen89} verified that vortex pairing is the most unstable two-dimensional secondary instability.   Strong stratification can inhibit vertical motion and suppress pairing \citep{Mashayek12b}. \citet{Mashayek12a} and \citet{Mashayek12b} 
show that three-dimensional secondary instabilities grow faster in high Reynolds number flows and can destroy the two-dimensional coherent structure required for vortex pairing. The critical Reynolds number at which pairing does not occur decreases with increasing Richardson number. \citet{Klaassen85}, \citet{Mashayek12a}, and \citet{Salehipour15} have shown that high Prandtl number can increase  the growth rate of some three-dimensional secondary instabilities, e.g. the secondary core instability.

However, in low to intermediate Reynolds number flows, which are applicable to some mixing layers and environmental flows \cite[e.g. see][]{breidenthal1981structure,rahmani14EFM}, vortex pairing is the dominant two-dimensional secondary instability. The pairing instability results from a coincident subharmonic of the most unstable wave number that forces neighboring KH billows to combine (pair). It can increase the vertical scale of motion and thickness of the shear layer \citep{Corcos84, Smyth93}. As a result, the effective Reynolds number is also increased. Since the amount of mixing and mixing efficiency are higher for higher Reynolds numbers in the mixing transition regime \citep{Rahmani14}, vortex pairing can enhance mixing and mixing efficiency.  The dominant three-dimensional secondary instability in this Reynolds number regime is the convective core instability \citep{Klaassen85, Caulfield00}.   \citet{Caulfield00} show that the growth rate of the convective core instability mainly comes from the mean shear, while the two-dimensional KH instability acts as a catalyst in the sense that it provides the flow on which the secondary instability grows.  The competition of vortex pairing and three-dimensional secondary instabilities determines whether vortex pairing occurs or not. This competition is dependent on the initial non-dimensional parameters, and also on the details of the initial perturbations \citep{Caulfield00, Metcalfe87}, e.g. the amplitudes of KH, the subharmonic components, and three-dimensional motions. 

Some researchers have studied the dependence of secondary instabilities on initial conditions in shear layers without density stratification, for example \citet{Patnaik76}, \citet{Ho82}, \citet{Ho84}, \citet{Metcalfe87}, \cite{hajj1993fundamental}, and \cite{husain1995experiments}. \citet{Patnaik76} show that shredding replaces pairing when the phase relationship between KH and the subharmonic modes is unfavourable for pairing. One vortex is strengthened and the other is weakened in that case. However, shredding is seldom observed in experiments due to the existence of ambient noise other than pure eigenfunctions of the Orr-Sommerfeld equation. \citet{Ho82} study the spreading rate of a spatially varied shear layer under different forcing. They show that without including the subharmonic mode in the initial perturbations pairing is significantly delayed.  \citet{Metcalfe87} demonstrate that vortex pairing can suppress the modal growth rate of a three-dimensional mode when the subharmonic mode reaches finite amplitude and the three-dimensional mode is small. However, this may only be valid for flows initialized by eigenfunctions of sufficient amplitudes. \cite{hajj1993fundamental} show that the growth of the subharmonic mode is maximum close to an optimal phase difference between the KH and the subharmonic mode and is suppressed at other phase differences. Similarly, \cite{husain1995experiments} demonstrate that at a phase difference unfavourable for pairing, and also for angles close to this phase difference, the growth rate of the subharmonic mode reduces significantly.  

  
Numerical investigations of shear instabilities in stratified flows have also found that vortex pairing depends on initial conditions, e.g. \citet{Klaassen89} and \citet{Smyth93}. \citet{Klaassen89} obtain the amplitude ratios of the first three harmonics with wavenumber $\frac{1}{2}\alpha_{kh}$, $\alpha_{kh}$, and $\frac{3}{2}\alpha_{kh}$, where $\alpha_{kh}$ is the wavenumber of the most unstable mode to the viscous Taylor-Goldstein (TG) equation \citep{Taylor31, Goldstein31}, in a two-wavelength domain from a numerical simulation perturbed by white noise. They demonstrate that pairing is delayed and the growth rate of the subharmonic mode is decreased if the subharmonic and the third modes are out of phase relative to KH instabilities. In general, the time of vortex pairing may be sensitive to the phase of the $\frac{3}{2}\alpha_{kh}$ mode if the subharmonic mode is out of phase with KH mode.  \citet{Smyth93} reached similar conclusion about the effect of the phase on pairing. 


Previous studies only considered the effect of initial conditions on pairing in two-dimensional simulations and mostly used eigenfunctions as initial perturbations, we extend these studies to examine the effects of phase difference between KH and subharmonic components in two- and three-dimensional flows with eigenfunction and random initial perturbations. 
Two-dimensional simulations are used to compare random perturbation simulations with eigenfunction perturbation simulations in terms of vortex pairing and  sensitivity of pairing to the phase difference between the KH and subharmonic mode. Three-dimensional simulations are used to investigate the effect of three-dimensional motions on pairing and mixing.

The paper is organized as follows. The numerical methods and diagnostic tools are described in section \ref{sec:methodology}. A simplified pairing mechanism is described in section \ref{sec:pair_mech}. Section \ref{sec:2D} describes the process of vortex pairing using the Lagrangian trajectory, the phase shift and the growth rate of the subharmonic mode in two-dimensional simulations. In section \ref{sec:3D}, three-dimensional results are compared with two-dimensional results to study the effect of three-dimensional motions and  mixing properties are compared in different simulations.

\section{Methodology}\label{sec:methodology}

\subsection{Mathematical Model}
The unperturbed background flow is a pure horizontal  stratified shear flow. The background velocity $\overline{U}$ and density $\overline{\rho}$ are hyperbolic tangent functions of vertical coordinate $z$, as first introduced by \citet{Hazel72},

\refstepcounter{equation}
$$
\bar{\rho} = -\frac{\Delta\rho}{2}\tanh\left(\frac{2z}{\delta_0} \right)
\quad \mbox{and\ } \quad
\overline{U} = \frac{\Delta U}{2} \tanh \left(\frac{2z}{h_0} \right),
\eqno{(\theequation{\mathit{a},\mathit{b}})}
$$
where $\Delta U$ and $\Delta \rho$ are the variations of velocity and density respectively, $\delta_0$ is the thickness of the density interface, and $h_0$ is the thickness of the velocity interface. Four non-dimensional parameters characterize the flows, i.e. the bulk Richardson number $J$, the Reynolds number $Re$, the Prandtl number $Pr$, and the scale ratio $R$ which are defined as
\begin{equation}
J = \frac{\Delta\rho gh_0}{\rho_0(\Delta U)^{2}}, \quad
Re = \frac{\Delta Uh_0}{\nu},\quad
Pr = \frac{\nu}{\kappa},\quad
R = \frac{h_0}{\delta_0},
\end{equation}
where $\kappa$ is molecular diffusivity, $\nu$ is kinetic viscosity, $\rho_0$ is a reference density.
In this study, $J=0.07$, $Re=1200$, $Pr=16$, $R=1$. The Reynolds number is small compared to that in many geophysical flows. The Prandtl number is slightly higher than the thermal Prandtl number of water at $20^{\circ}$C which is about 7.  The flow is susceptible to Kelvin-Helmholtz instabilities for this combination of $J$ and $R$ \citep[see][for a review of instability types ]{Smyth03}.

We assume the fluid is incompressible and apply the Boussinessq approximation for small density difference, so the governing equations for the system are
\begin{equation}
\nabla\cdot\boldsymbol{u} = 0, \label{eqn:1}
\end{equation}

\begin{equation}
\frac{D\boldsymbol{u}}{Dt} = -\frac{1}{\rho_0}\nabla p - \frac{\rho}{\rho_0}g\hat{\boldsymbol{k}} + \nu\nabla^{2}\boldsymbol{u}, \label{eqn:2}
\end{equation}

\begin{equation}
\frac{D\rho}{Dt} = \kappa\nabla^{2}\rho, \label{eqn:3}
\end{equation}
where $u$ and $p$ are the fluid's velocity and pressure respectively and $\hat{\boldsymbol{k}}$ is the unit vertical vector. $D/Dt$ is the  material derivative and $g$ is the gravitational acceleration.

\subsection{Direct Numerical Simulations}\label{subsec:dns}
\begin{table}
	\begin{center}
		\def~{\hphantom{0}}
		\begin{ruledtabular}\begin{tabular}{ccccccccc}
			$Re$ & $Pr$ & $J$ & $L_x/h_0$ & $L_y/h_0$ & $L_z/h_0$ & $N_x$ & $N_y$ & $N_z$  \\
			1200 & 16 & 0.07 & 14.43 & 7.22 & 15 & 320 & 160 & 320 \\
		\end{tabular} \end{ruledtabular}
		\caption{Numerical parameters for all the simulations. The number of grid points  is for the velocity field and is half that of the density field.}
		\label{tab:parameters}
	\end{center}
\end{table}

The governing equation (\ref{eqn:1}), (\ref{eqn:2}), and (\ref{eqn:3}) are solved by a pseudo-spectral code developed by \citet{Winters04} and later improved by \citet{Smyth05}. The code employs a third order Adams-Bashforth time stepping scheme. Boundary conditions are horizontally periodic and vertically free slip and no flux for our simulations.

The domain length $L_x$ is set  to two wavelengths of the most unstable mode to allow vortex pairing. The spanwise width of the domain $Ly$ for the three-dimensional simulations is one wavelength of the primary KH instability, which is at least six wavelengths of most unstable spanwise mode \citep[for the most unstable spanwise wavenumber see][]{Klaassen89}. The domain height is $15h_0$ which is sufficient to remove the effects of the top and bottom boundaries on pairing. The numerical details are summarized in table \ref{tab:parameters}.

The resolution of DNS are typically determined by the Kolmogrov scale, $L_k=(\nu^3/\varepsilon ')^{1/4}$, in homogeneous fluids where $\varepsilon '$  is the viscous dissipation rate of turbulent kinetic energy. \citet{Moin98} suggest that the grid spacing in DNS should be $\textit{O}(L_k).$ In stratified flows with $Pr>1$, the smallest  scale that need to be resolved it $\textit{O}(L_B)$ where $L_B$ is the Batchelor scale \citep{Batchelor59} and $L_B=L_k/{\sqrt{Pr}}$. In our  simulations, $\Delta z/{L_B}$ is always less than 4.0 and $\Delta z/{L_K}$ is always less than 2.0 (grid spacing of the density field is half of that of the velocity field.). The dissipation rate $\varepsilon '$ used to calculate $L_K$ is averaged within $L_z/2-h_0/2<z<L_z/2+h_0/2$ where turbulence is the most energetic. Although previous studies have used  finer resolutions, e.g. \citet{Smyth03} and \citet{Rahmani14}, more recent studies, e.g. \citet{Salehipour15} have used  similar resolutions. We also ran a resolution test for simulation $R\frac{\pi}{2}3D$ (table 2) with a double resolution. The final amount of mixing in the simulation with the finer resolution changed by less than $0.1\% $.

We ran two sets of simulations to study the effect of initial perturbations on vortex pairing and mixing. One set is perturbed by random perturbations, where three simulations are performed in both two and three dimensions. The energy of initial random perturbations projected on each two-dimensional Fourier component is almost the same. The other set is perturbed by the eigenfunctions to the TG equation of the KH and the subharmonic modes with the same amount of kinetic energy of the KH and the subharmonic mode (defined in equation (\ref{eqn:spec_energy})) as in random perturbation simulations. The eigenfunctions are obtained by solving TG equation using a second-order finite difference method. The eigenfunction simulations are performed in two dimensions only. The simulations are listed in table \ref{tab:times} with key resultant times.

For random perturbation simulations, inherently three dimensional random perturbations of $u'$ and $w'$ are added to the background flow to excite instabilities in three-dimensional simulations. They are given by the following equations,
 \begin{equation}
 u' =a r_u(x,y,z)\frac{\Delta U}{2}\left(1-\left| \tanh\left( \frac{2z}{h_0} \right)\right| \right),\label{eqn:u'}
 \end{equation}
 
 \begin{equation}
 w' = ar_w(x,y,z)\frac{\Delta U}{2}\left(1-\left| \tanh\left( \frac{2z}{h_0} \right)\right| \right),
 \label{eqn:w'}
 \end{equation}
 where $r_u$ and $r_w$ are random numbers between -1 and 1, and $a$ sets the maximum amplitude of perturbations.
In the present study, $a=0.1$, as in the simulations of \citet{Smyth03} and \citet{Carpenter10}, and small enough for perturbations to grow linearly initially. 
The initial conditions in two-dimensional simulations are spanwise averaged values of those in corresponding three-dimensional simulations.

We define the phase of each wavenumber component in terms of two-dimensional vertical velocity $w_{2d}$ (defined in equation \ref{eqn:u_2d} (b)) , i.e.,
\begin{equation}
\theta_k=\frac{\pi}{2} + arg \left\{
\hat{w}_{2d,k}\left(z=\frac{L_z}{2}\right)\right\},\label{eqn:phase}
\end{equation}
where $arg$ is the argument or phase of a complex number, $\frac{\pi}{2}$ is added to make the vertical velocity $\Re\{\hat{w}_{2d,k}e^{i2k\pi x/L_x}\}$ of a specific mode a sine wave, and
$\hat{w}_{2d,k}$ is the $k$th Fourier component of spanwise averaged vertical velocity $w_{2d}$. \citet{Klaassen89} and \citet{Smyth93} use a similar definition in terms of streamfunction. We use ``component" to denote the Fourier component and ``mode" to denote eigenfunction of a specific wavenumber throughout the paper. When a component becomes approximately modal after a non-modal growth, we call it a mode. Note that $k=2$ corresponds to the KH component and $k=1$ corresponds to the subharmonic component, i.e. a wavelength equal to the domain length, $L_x$. 

Initially each Fourier component experiences a non-modal growth and the phase of every component defined in equation  (\ref{eqn:phase}) changes. When the subharmonic component becomes approximately modal, i.e. identical to the eigenfunction of the TG equation (this occurs around non-dimensional time $t\Delta U/h_0=24$ for the random perturbation simulations), the phase becomes almost constant for some time until non-linear effects become important.   We define $\theta_{sub}$ as the phase of the subharmonic component relative to the KH component and  $\theta_{sub}^M$ as the value of $\theta_{sub}$ when the subharmonic component  becomes modal (defined in section \ref{sec:modality}). We examine the effects of phase difference between the KH and subharmonic mode by considering three different phases in our random perturbation simulations. 
We create these phase differences by first generating a random perturbation using equation (\ref{eqn:u'}) and (\ref{eqn:w'}) and calculating its phase $\theta_{sub}^M$, and then multiplying the coefficient of the subharmonic term of this perturbation  by $e^{i\Delta\phi}$, where $\Delta\phi$ is the desired phase shift. These random simulations are designated by $R$ followed by the approximate modal phase $\theta_{sub}^M$, and $2D$ or $3D$ depending on the number of dimensions of the simulation. For example, $\theta_{sub}^M$ is approximately $0$, $-\frac{\pi}{4}$, and $-\frac{\pi}{2}$ respectively  in three-dimensional simulations $R03D$, $R{\frac{\pi}{4}}3D$, and $R{\frac{\pi}{2}}3D$ (exact phase values are listed in table \ref{tab:times}). It will be explained in section \ref{sec:discussion} that the sign of the phase is not important.

Simulations perturbed by eigenfunctions are named by the same procedure, but the first letter is $E$, indicating that they are perturbed by eigenfunctions. For these eigenfunction perturbed simulations, the phase of the subharmonic mode does not change initially and $\theta_{sub} ^M$ is equal to the initial phase value. $\theta_{sub}^M = -\frac{\pi}{2}$ for $E{\frac{\pi}{2}}2D$ and $\theta_{sub}^M = 0$ for $E02D$.  

\begin{table}
	\begin{center}
	\begin{minipage}{1.0\linewidth} 
	                \begin{center}
	                 \begin{ruledtabular} \begin{tabular}{c cccccc c cc}
	                 
		& \multicolumn{6}{c}{Random perturbations} & &\multicolumn{2}{c}{Eigenfunctions} \\
		\cline{2-7} \cline{9-10}
			Run     & $R02D$ &   $ R{\frac{\pi}{4}}2D$    &  $R{\frac{\pi}{2}}2D$  &  $R03D$   &  $R{\frac{\pi}{4}}3D$  & $R{\frac{\pi}{2}}3D$ & &$E02D$  &   $E{\frac{\pi}{2}}2D$  \\\cline{2-4} \cline{5-7} \cline{9-10}
			$ \theta_{sub}^M/\pi $  &  0.04  & - 0.21 & -0.46 & 0.05  & -0.21 & -0.48 & &0 & 0.5\\
			$t_M$ & 24 & 24 & 24 & 24 & 24 & 24 & & 0 & 0 \\ 
			
			$t_{kh}$     & 80    &     82    &   85     & 80   & 81  & 84    && 81 & 87    \\
			$t_{sub}$   & 106   &    110   & 146   &  107  &113  &129 && 104    & 249    \\
			$t_p$         & 108   &    112   & 148  & 109  &113   & --     && 107     & 249    \\
			$t_{3d}$    &   --     &    --      & --     &  143   &146  & 123  & & --      & --     \\

		\end{tabular} \end{ruledtabular}
						
		\begin{minipage}{1.0\linewidth}
        		\begin{tabular}{clr} 
		 Parameter& Definition&Reference Figure\\
		$ \theta_{sub}^M$&  phase difference between  the primary KH & Fig. 4(a)\\
		  &   and  subharmonic mode & \\
	                 $t_M$ &onset of modal growth, time when $r=0.99$& Fig. 3\\
		$t_{KH}$& first peak in kinetic energy of the primary KH & Fig. 6\\
		$t_{sub}$& global peak in the kinetic energy of the subharmonic& Fig. 6\\
		$t_p$ & when pairing vortices initially cross&Fig. 4(b)\\
		$t_{3d}$& first peak in the 3D kinetic energy& Fig. 6\\  
		 \end{tabular} 
	                   \end{minipage} \\		
		\end{center}
	\end{minipage}
		
		\caption{Definition of phase and important times and list of simulations. The simulations beginning with $E$ are perturbed by eigenfunctions, while those beginning with the letter $R$ are perturbed by random perturbations. The phase difference (0, $\pi/4$ or $\pi/2$) and the dimensions of each simulation appear in the name of the simulation. We also performed three-dimensional random perturbation simulations with phase differences $-0.125\pi$, $-0.375\pi$ and $-0.45\pi$ (not listed in the table) for mixing quantifications.}
		\label{tab:times} 
	\end{center}
\end{table}

\subsection{Diagnostic tools}\label{sec:diag}
Following \citet{Caulfield00},  the velocity is decomposed into three parts, i.e.,

\refstepcounter{equation}
$$
\overline{\boldsymbol{u}} = {\langle\boldsymbol{u}\rangle}_{xy},\quad
\boldsymbol{u}_{2d} = {\langle\boldsymbol{u}\rangle}_y - 
{\langle\boldsymbol{u}\rangle}_{xy} \label{eqn:u_2d},\quad
\boldsymbol{u}_{3d} = \boldsymbol{u} - \overline{\boldsymbol{u}}
- \boldsymbol{u}_{2d},
\eqno(\theequation{\mathit{a}\mbox{--}\mathit{c}}) \label{eqn:vel_component}
$$
where the subscripts indicate averaging over that direction.
Given these definitions, the total kinetic energy $K$  is defined as 
\begin{equation}
K = \frac{{\langle\boldsymbol{u}\cdot\boldsymbol{u}\rangle}_{xyz}}{2\rho_0 \Delta U^2},
\end{equation}
where $\rho_0 \Delta U^2$ is used for non-dimensionalization,  
and can be partitioned into three parts $\overline{K}$, $K_{2d}$, $K_{3d}$, i.e.,
\begin{equation}
K = \overline{K} + K_{2d} + K_{3d},
\end{equation}
where

\refstepcounter{equation}
$$
\overline{K} = \frac{{\langle\overline{\boldsymbol{u}}\cdot\overline{\boldsymbol{u}}\rangle}_z}{2\rho_0 \Delta U^2},\quad
K_{2d} = \frac{{\langle{\boldsymbol{u}}_{2d}\cdot{\boldsymbol{u}}_{2d}\rangle}_{xz}}{2\rho_0 \Delta U^2},\quad
K_{3d} = \frac{{\langle{\boldsymbol{u}}_{3d}\cdot{\boldsymbol{u}}_{3d}\rangle}_{xyz}}{2\rho_0 \Delta U^2},\quad
\eqno{(\theequation{\mathit{a}\mbox{--}\mathit{c}})}.
$$

Fourier transforms are applied to $u_{2d}$ and $w_{2d}$ in order to identify the contribution of each wavenumber component  $K_{2d}$, so that the kinetic energy of the $k$th component is
\begin{equation}
K_{2d,k}= \frac{{\langle |\hat{u}_{2d,k}|^2 + |\hat{w}_{2d,k}|^2 \rangle}_{z}}{\rho_0 \Delta U^2},\quad k\geqslant 1
\label{eqn:spec_energy}
\end{equation}
where $\hat{u}_{2d,k}$ and $\hat{w}_{2d,k}$ are the Fourier components of $u _{2d}$ and $w_{2d}$ of wavenumber $2\pi k/L_x$. Hence, 
\begin{equation}
K_{2d} = \sum_{k=1}^{\frac{N_x}{2}}K_{2d,k}.
\end{equation}
Note that $k=1$ corresponds to the subharmonic component and we denote it as $K_{sub}$. $k=2$ corresponds to the KH instability and we denote it as $K_{kh}$. These two components of the kinetic energy characterize the kinetic energy of the  subharmonic and  primary components. 

We follow the framework in \citet{Winters95} to study mixing. The potential energy is then defined as,
\begin{equation}
P =\frac{g{\langle\rho z\rangle}_{xyz}}{\rho_0\Delta U^2}.
\end{equation}
Potential energy $P$ is partitioned into background potential energy $P_b$ and available potential energy $P_a$ defined as
\begin{equation}
P_b = \frac{g{\langle\rho z_b\rangle}_{xyz}}{\rho_0\Delta U^2},\quad
P_a = P - P_b,
\end{equation}
where $z_b$ is the location of fluid parcels after being re-arranged into a statically stable state \citep[see][]{Winters95}. 
 Available potential energy characterizes the energy that can be exchanged between potential energy and kinetic energy, while the increase in background potential energy quantifies irreversible mixing in a closed system. The amount of mixing caused by the fluid's motion is
\begin{equation}
M = \Delta P_b - D  \equiv \int\phi_M dt,
\end{equation}
where $\phi_M$ is defined as the rate of mixing and $D $ is the mixing caused by molecular diffusion in quiescent fluid and calculated by
\begin{equation}
D  =  \frac{-\kappa g ( \bar{\rho}|_{z=L_z}-\bar{\rho}|_{z=0})t}{L_z} \frac{1}{\rho_0\Delta U^2}.\label{eqn:phi_i}
\end{equation}
During the whole process, $D$ grows approximately linearly.
The instantaneous mixing  $\phi_M$ is always positive and varies over time. Cumulative mixing efficiency \citep{Caulfield00} is used as a measure of overall mixing properties in this study. It is defined as 
\begin{equation}
E_c = \frac{\displaystyle{\int_{t_{3d}}^{t_f} \phi_M dt}}{\displaystyle{\int_{t_{3d}}^{t_f} \phi_M dt + \int_{t_{3d}}^{t_f} \varepsilon dt}},\label{eqn:Ec}
\end{equation}
where $t_{3d}$ is the time when $K_{3d}$ reaches its maximum and $t_f$  is defined as the time when buoyancy Reynolds number $Re_b=\varepsilon '/\nu{\langle N^2\rangle}_z$ first drops below 20 after $t_{3d}$. 
This period is chosen as previous investigations show that turbulence is active only when $Re_b>20$ \citep{Smyth00}. By choosing $t>t_{3d}$, we remove the two-dimensional mixing because mixing caused by two-dimensional overturns is process dependent \citep{Salehipour16,Mashayek17}  and specifically depends on initial perturbation. Hereafter, time is non-dimensionalized by $h_0/\Delta U$ and we refer $t$ as the dimensionless time.

\begin{figure}
	\centerline{\includegraphics[width = 0.8\textwidth]{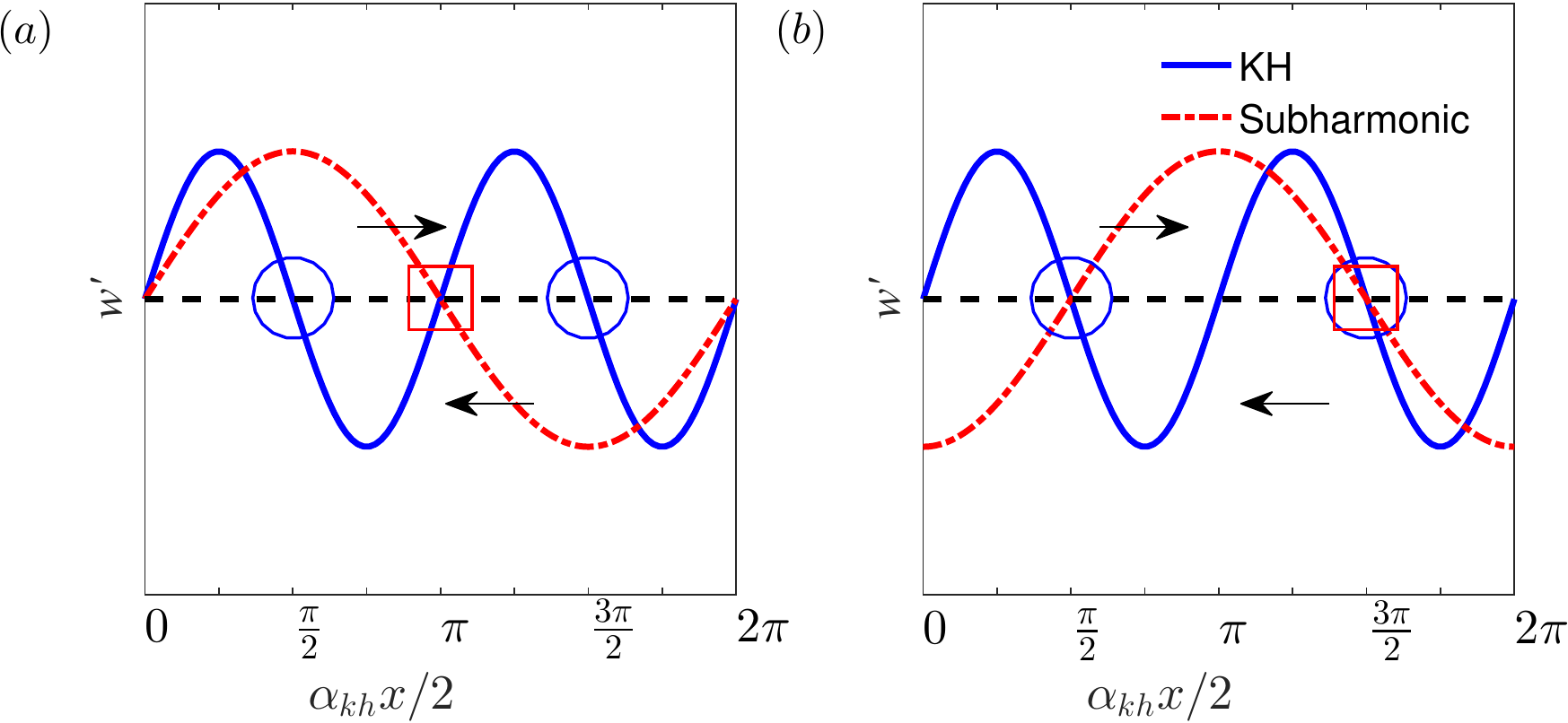}}
	\caption{Conceptual drawing of vortex pairing demonstrating the effect of phase of the subharmonic mode. (\textit{a}) $\theta_{sub}=0$ (\textit{b}) $\theta_{sub}=-\frac{\pi}{2}$. The blue solid line is KH mode and the red dash-dotted line is the subharmonic mode. Blue circles denote the location of KH vortex centers and red squares are vortex centers associated with the subharmonic mode. The two arrows show the directions of the mean flow.} 
	\label{fig:cartoon}
\end{figure}

\section{Pairing mechanism} \label{sec:pair_mech}
The pairing process and the importance of the phase of the subharmonic are illustrated in figure \ref{fig:cartoon}. In figure \ref{fig:cartoon} ($a$), the subharmonic mode displaces the left KH billow upward and the right KH billow downward. The two KH billows are then advected toward each other by the mean flow, cross each other, and merge into one larger billow. This is the optimal phase for pairing. In figure \ref{fig:cartoon}  ($b$), the phase of the subharmonic mode is $\theta_{sub} = -\frac{\pi}{2}$ and two KH core centers are at the nodes of the subharmonic mode. This is called the ``shredding mode'' in \citet{Patnaik76} and the ``draining mode" in the discussions by  \citet{Klaassen89} and \citet{Smyth93}.  In this case, one KH vortex (the right one in figure \ref{fig:cartoon} $b$) is strengthened by the subharmonic mode and the other KH vortex (the left one in figure \ref{fig:cartoon} $b$ ) is weakened by the straining field of the subharmonic mode. For example, in figure \ref{fig:cartoon} ($b$), the right vortex will be stronger than the left one. 

\begin{figure}
	\centerline{\includegraphics[width = 1.3\textwidth]{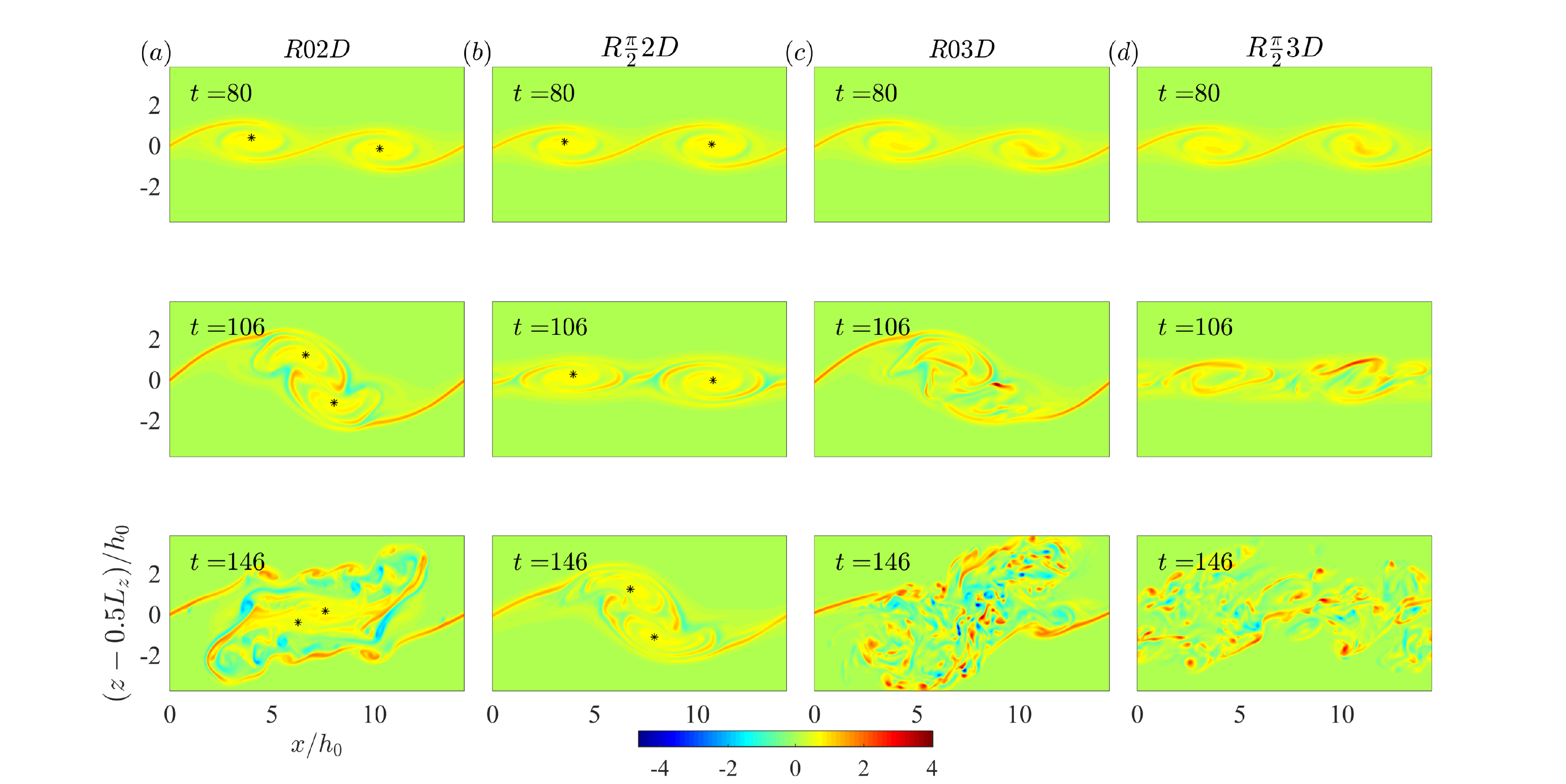}}
	\caption{Non-dimensional span-wise vorticity $(u_z - w_x)h_0/\Delta U$  of two-dimensional simulations $R02D$ (phase of the subharmonic mode is $\theta_{sub} ^M \approx 0$) and $R{\frac{\pi}{2}}2D$ (phase of the subharmonic mode is $\theta _{sub}^M \approx -\displaystyle\frac{\pi}{2}$) and their corresponding three-dimensional simulations $R03D$, and $R{\frac{\pi}{2}}3D$. The snapshots of three-dimensional simulations are plotted at $\displaystyle y=\frac{L_y}{2}$. Pairing is delayed in two-dimensional simulation $R{\frac{\pi}{2}}2D$ but completely eliminated in the three-dimensional simulation $R{\frac{\pi}{2}}3D$. Black stars are fluid particles located at vortex centres at $t=30$.}
	\label{fig:vort_snap}
\end{figure}

Resultant KH billows with and without pairing are illustrated in the vorticity snapshots from DNS in figure \ref{fig:vort_snap}. At $t=106$, the simulation with the phase of subharmonic mode $\theta_{sub} = 0$,  $R02D$, is undergoing a vortex merging, while the simulation with $\theta_{sub} = -\frac{\pi}{2}$, $\theta_{sub} = 0$,  $R\dfrac{\pi}{2}2D$, exhibits a draining mode. In $R\dfrac{\pi}{2}2D$, the pairing mode eventually grows and surpasses the KH mode. During this adjustment, the phase of the subharmonic mode shifts toward 0. We discuss this pairing process in section \ref{sec:2D}. In three-dimensional simulations, the growth of three-dimensional motions disintegrates the two-dimensional structure of the billows and can inhibit the merging of the billows, see simulations $R03D$ and $R\dfrac{\pi}{2}3D$ at $t=146$. We discuss these effects in section \ref{sec:3D}.

\section{Two-dimensional aspects of  pairing}\label{sec:2D}
In this section, we examine the 2D pairing process focusing on comparing pairing in flows perturbed by eigenfunctions with flows perturbed by random perturbations. Besides the traditional phase and growth rate analysis, we characterize the degree of modality quantitatively and use Lagrangian trajectories to aid in the interpretation of the Fourier decomposition.

\begin{figure}
	\centerline{\includegraphics[width = 0.6\textwidth]{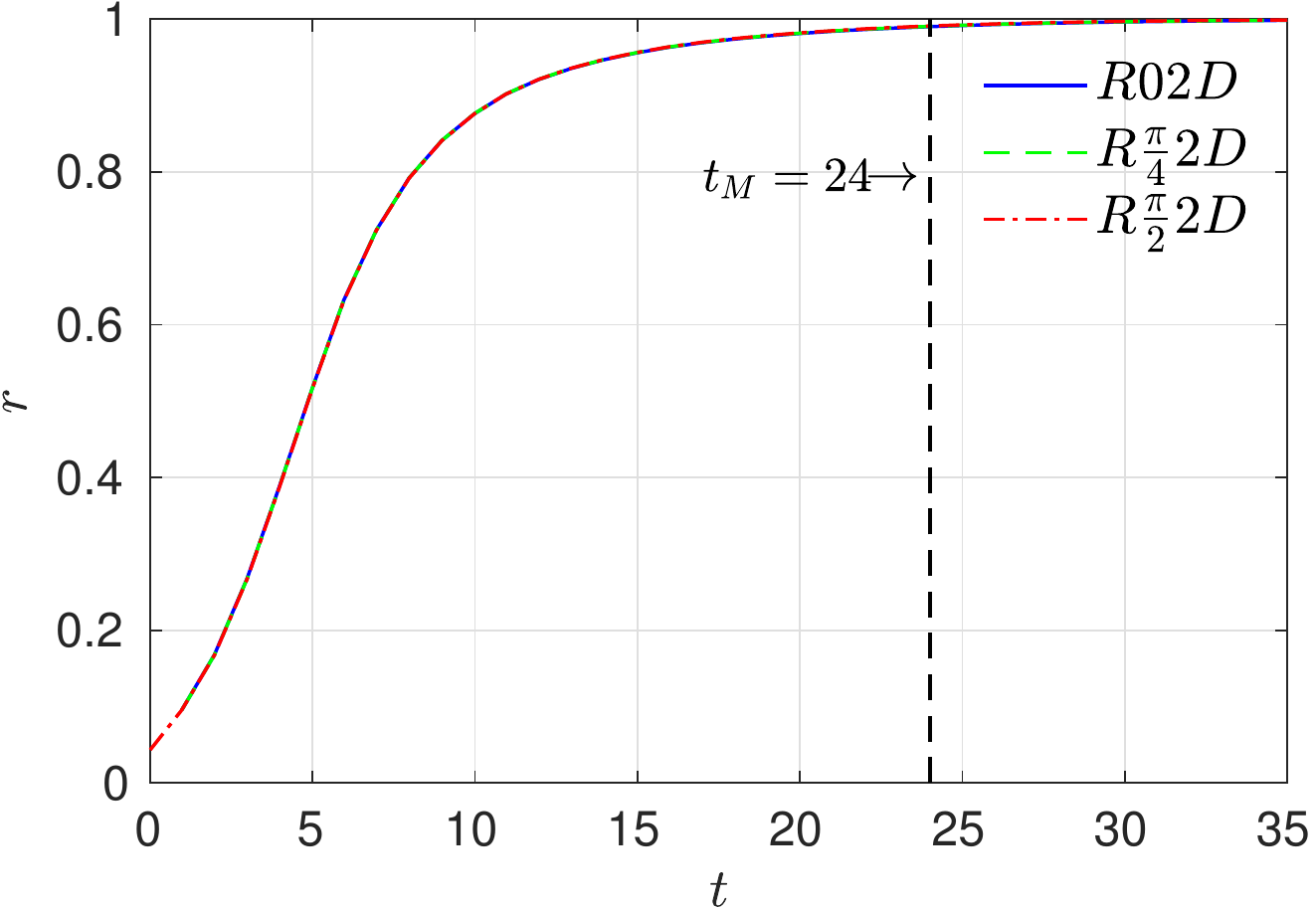}}
	\caption{Evolution of $r$ for  random perturbation simulations $R02D$, $R{\frac{\pi}{4}}2D$, and $R{\frac{\pi}{2}}2D$.} 
	\label{fig:beta}
\end{figure}

\subsection{Degree of Modality}\label{sec:modality}
We use the cosine of Hermitian angle \citep{Scharnhorst01} between the subharmonic component $\hat{w}_{2d,sub}$ and the initial eigenfunction of the subharmonic mode $\hat{w}_{eig,sub}$ to quantify the degree of modality,

\begin{equation}
r(t)  =  \frac{ |(\hat{w}_{2d,sub}, \hat{w}_{eig,sub}) | } {| \hat{w}_{2d,sub}| | \hat{w}_{eig,sub}|}
\label{eqn:nonmodal_measure}
\end{equation}
where  $(\cdot,\cdot )$ denotes the standard scalar product for complex vectors and $||$ denotes the amplitude of a complex number. $r(t)$ is the ratio between the length of the orthogonal projection of the subharmonic component  onto the the eigenfunction to the length of itself. It is always between 0 and 1, and  equal to 1 only when the subharmonic component of the random perturbation is identical to the eigenfunction.  

Figure \ref{fig:beta} shows the evolution of $r$ for the three random perturbation simulations $R02D$, $R{\frac{\pi}{4}}2D$, and $R{\frac{\pi}{2}}2D$. Initially, $r$ is small because the subharmonic component of the initial random perturbations is significantly different from the eigenfunction. As the subharmonic component evolves to the eigenfunction, $r$ increases to 1. We define the time required for the subharmonic component to become modal, $t_{M}$, as the time when $r$ first exceeds 0.99, which is  $t=24$ for these three simulations. Before $t=24$, the three simulations appear identical in figure \ref{fig:beta} because  the subharmonic component is evolving linearly, i.e., non-linear interaction of different components is negligible.

\subsection{Phase evolution}
\begin{figure}
	\centerline{\includegraphics[width=\textwidth]{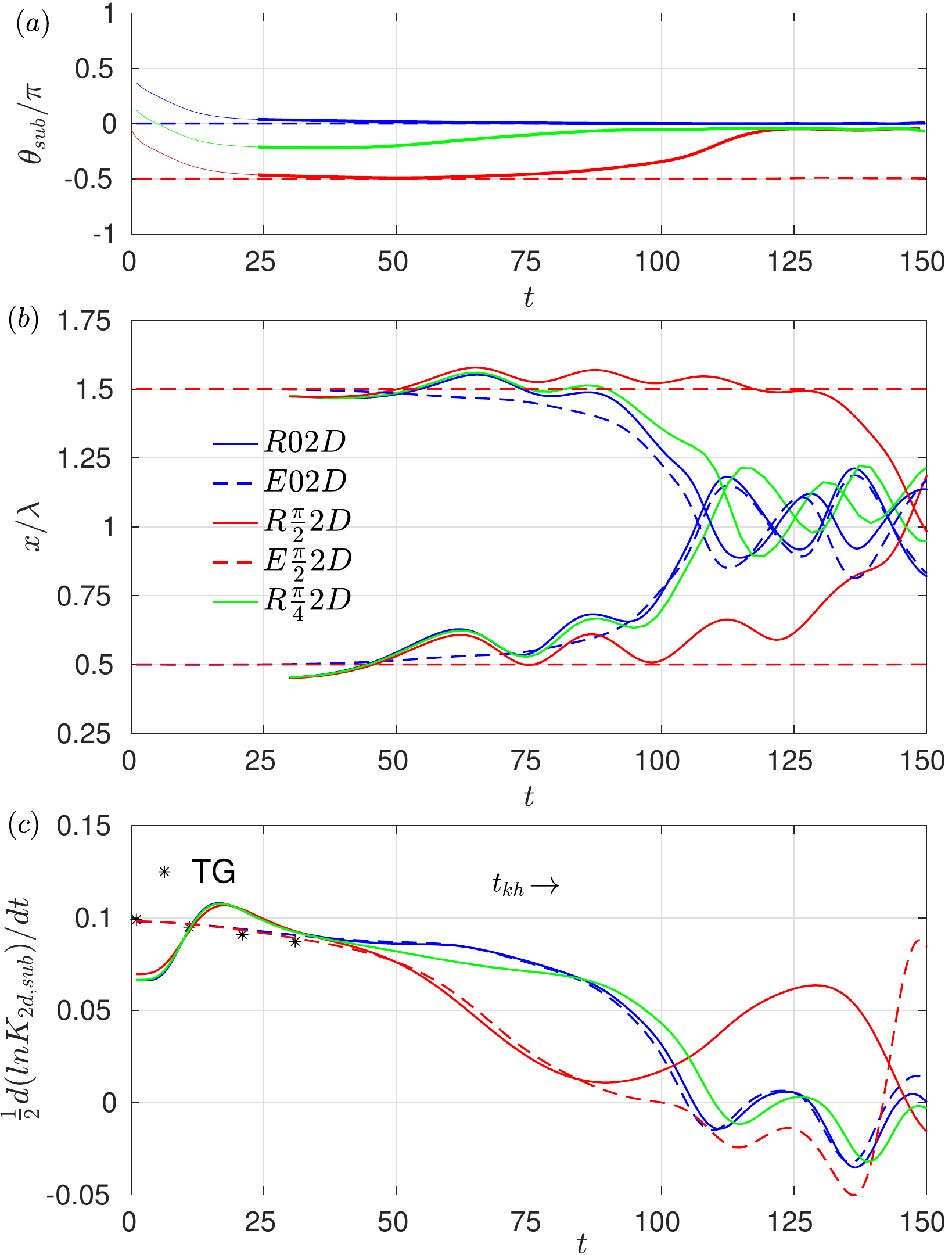}}
	\caption{($a$) Evolution of the phase of the subharmonic component. The phase before $t_M$ is shown as thin lines. The phase shift in simulation $E{\frac{\pi}{2}}2D$ begins after $t=150$ and is not shown in this figure. ($b$) $x$ coordinates of two fluid particles located at the two vortex centers at $t=30$ for random perturbation simulations and at $L_x/4$ and $3L_x/4$ for eigenfunction perturbation simulations. Pairing occurs at $t=249$ for simulation $E{\frac{\pi}{2}}2D$ and is not shown in this figure. ($c$) Growth rate of the subharmonic mode. The vertical dashed line indicates the saturation time of KH instabilities in simulation $R{\frac{\pi}{4}}2D$. The stars labelled as TG indicate the growth rate calculated using the TG equation with the time-dependent mean flow.} \label{fig:growth_rate}
\end{figure}

The phases for the five two-dimensional simulations perturbed by random perturbations and the eigenfunctions are plotted in figure \ref{fig:growth_rate} ($a$). Initially, in the eigenfunction perturbation simulations, the phase does not change. In the three random perturbation simulations, before $t=24$, the phases change because of non-modal growth. The pre-modal phase (before $t_M$) in the random perturbation simulations is therefore shown as a thin line. Between $t=24$ and $t=50$, the phases stay almost constant. During this period, the phases are approximately $0, -\frac{\pi}{4}, -\frac{\pi}{2}$ for $R02D$, $R{\frac{\pi}{4}}2D$, and $R{\frac{\pi}{2}}2D$, respectively (see table \ref{tab:times}). After $t=50$, the phases of the subharmonic mode in simulations $R{\frac{\pi}{4}}2D$ and $R{\frac{\pi}{2}}2D$ shift toward 0, which is similar to the results of \citet{Klaassen89}.
For the eigenfunction simulation $E{\frac{\pi}{2}}2D$, the phase begins to shift after  $t=200$ and is not shown in the figure. We ran a supplementary simulation perturbed by the KH, the subharmonic, and a third mode of wavenumber $\frac{3}{2}\alpha_{kh}$ and found that the time of pairing is greatly reduced compared to the $E{\frac{\pi}{2}}2D$ simulation. This result is consistent with the earlier phase shift found in the three-mode simulations of \citet{Klaassen89}.

\begin{figure}
	\centerline{\includegraphics[width = 0.6\textwidth]{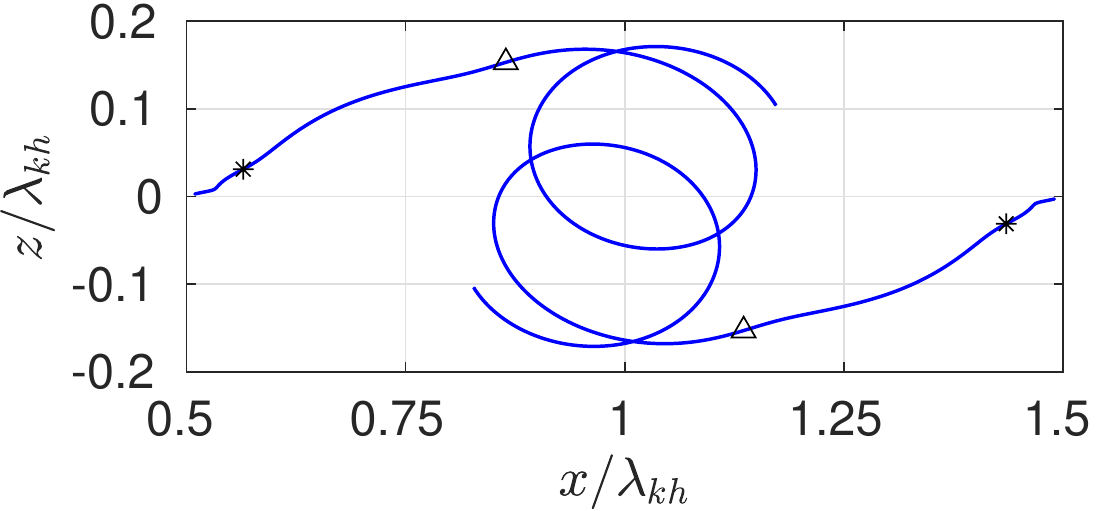}}
	\caption{The trajectories of two fluid particles between $t=45$ and $t=135$. The stars indicate $t_{kh}$ and the triangles indicate $t_{sub}$} 
	\label{fig:x-z}
\end{figure}

\subsection{Trajectories of pairing KH billows}
To characterize the trajectories of KH billows during pairing, fluid particles at the two inflection points of the  contour $\rho = \rho_0$ that correspond to the KH vortex centres are tracked. In the randomly perturbed simulations the vortex centres are identified at $t=30$, the earliest time when the KH vortices are clearly identifiable. In the case of the eigenfunction simulations the vortex centres are initially at $\frac{L_x}{4}$ and $\frac{L_x}{2}$. 
Figure \ref{fig:vort_snap} shows the evolution of the vorticity field with the two fluid particles shown as black stars. As the figure shows, these two fluid particles approximately represent the vortex centres until small-scale motions prevail, e.g. at $t = 146$ for simulation $R02D$. As an example,  figure \ref{fig:x-z} shows the trajectories in the eigenfunction perturbation simulation $E02D$ with the optimal phase.  The trajectories of the two particles are well organized and symmetric about the domain centre because of the symmetry of the initial conditions. From $t_{kh}$, defined as the time when $K_{kh}$ reaches its first maximum (the global maximum is caused by vortex pairing),  to $t_{sub}$, defined as the time when $K_{sub}$ reaches its global maximum, the KH billows undergo most of their vertical displacement. After this time, the billows cross over each other merging into the subharmonic billow, while the  two vortex centres rotate around the domain centre. The KH billow originally at the crest of the subharmonic remains, on average, elevated above the KH billow originally at the trough of the subharmonic. As will be shown in section \ref{sec:3D}, during this first orbit ($t = 107$ to $t = 131$) three-dimensional motions become important.

Now we return to figure \ref{fig:growth_rate} ($b$) which shows the temporal variability of the horizontal coordinates of the two fluid particles. After approximately $t=75$, the two fluid particles quickly converge in $x$ for the two optimal phase simulations $E02D$ and $R02D$. We defined the time of pairing, $t_p$, as the time when the horizontal distance first becomes zero and listed in table \ref{tab:times} (for three-dimensional simulations, $t_p$ is obtained by averaging two sets of trajectories each composed of 21 fluid particles spread over the spanwise direction). The two vortices become closer and merge in simulation $E{\frac{\pi}{2}}2D$ at $t=249$ (not shown in the figure). Unlike the results of the eigenfunction simulations, in the random perturbation simulations there is  an oscillation  of the fluid particles before $t_{sub}$ and $t_p$ because of the existence of modes other than the KH and subharmonic modes. For these three random perturbation simulations, the horizontal distance between the two fluid particles is always the smallest for $R02D$ and largest for $R{\frac{\pi}{2}}2D$. Also, pairing occurs first in $R02D$ and last in $R{\frac{\pi}{2}}2D$, so $t_p^{R02D}<t_p^{R{\frac{\pi}{4}}2D}<t_p^{R{\frac{\pi}{2}}2D}$.  Relating figure 4 ($b$) with the phase evolution in figure \ref{fig:growth_rate} ($a$), we find that the two fluid particles begin to move together (i.e. pair) only once the phase is approximately optimal. The pairing therefore occurs earliest if the subharmonic is in phase and latest if it is out of phase similar to the previous studies  of \citet{Klaassen89}, \citet{Smyth93}, \citet{hajj1993fundamental} and \citet{husain1995experiments}. Comparison between $t_p^{R{\frac{\pi}{4}}2D}$ and $t_p^{R02D}$ indicates that if the subharmonic mode is not close to $\pm \frac{\pi}{2}$, the difference in pairing is small, as also observed by \citet{husain1995experiments}. Also, the time of pairing for simulation $R02D$ is close to $E02D$, but the time of pairing for simulation $R{\frac{\pi}{2}}2D$ is much earlier than in simulation $E{\frac{\pi}{2}}2D$.


\subsection{Growth rate}
Figure \ref{fig:growth_rate} ($c$) shows the growth rate of the subharmonic component for the three random perturbation simulations and the two eigenfunction simulations.  Initially, the growth rates of the subharmonic mode for the two eigenfunction simulations ($E02D$ and $E{\frac{\pi}{2}}2D$) are the same and decline as the shear layer diffuses. The estimated growth rate (labeled with TG) using the TG equation and the varying mean flow  has the same decreasing trend as the growth rates based on $K_{sub}$. For the random perturbation simulations ($R02D$, $R{\frac{\pi}{4}}2D$, and $R{\frac{\pi}{2}}2D$), the growth rate during the non-modal stage of growth can be either smaller or larger than the modal growth rate as found by \citet{Guha14}. For all five simulations,  the growth rate is independent of the phase  in the initial linear stage of growth, i.e. before around $t=45$. 

After $t=45$,  non-linear effects and the phase become important. After $t=45$ and before $t_{kh}$, the growth rates in the late pairing simulations ($E{\frac{\pi}{2}}2D$ and $R{\frac{\pi}{2}}2D$) drop compared to the other simulations, in agreement with  \citet{Klaassen89}. However, we find that the growth rate in the $R{\frac{\pi}{4}}2D$ simulation stays closer to the optimal phase simulations $R02D$ and $E02D$. The growth rates in simulations $R02D$ and $E02D$ are almost the same and the growth rates in simulations $R{\frac{\pi}{2}}2D$ and $E{\frac{\pi}{2}}2D$ are almost the same. This suggests that if the phase and amplitude of the subharmonic and KH are the same for an eigenfunction perturbation and a random perturbation, the growth rate of the subharmonic mode is the same before saturation of KH instabilities. In other words,  the existence of the other components in initial perturbations and initial non-modal growth have negligible effects on the growth rate during this non-linear growth stage.

When the KH instability reaches its maximum amplitude ($t_{kh}$) the phase is close to optimal in $R02D$, $R{\frac{\pi}{4}}2D$, and $E02D$. The growth rates then quickly decrease to zero. In these three simulations, the first zero crossing of the growth rate is close to $t_p$ and denotes the saturation of the subharmonic mode, i.e. the global maximum of $K_{sub}$. 
In $R{\frac{\pi}{2}}2D$, after $t_{kh}$ the growth rate begins to increase along with the phase shifting toward the optimal value (see figure \ref{fig:growth_rate} ($a$)). In this simulation, the saturation of the subharmonic mode occurs   at $t=146$. In $E{\frac{\pi}{2}}2D$, the phase remains at $-\frac{\pi}{2}$ and the growth rate continues to decrease. Unlike the other simulations, the growth rate crosses zero before saturation of the subharmonic mode (at $t=249$). 
 Comparison between simulations $R{\frac{\pi}{2}}2D$ and $E{\frac{\pi}{2}}2D$ shows that the growth rate is sensitive to the initial structure of the subharmonic component or the existence of the other components in initial conditions if the phase is close to $- \frac{\pi}{2}$.
 
In table \ref{tab:times}, the saturation times of KH and the subharmonic mode and time of pairing ($t_{kh}, t_{sub}$, and $t_{p}$) are summarized. Pairing occurs first in simulation $E02D$, second in $R02D$, third in $R{\frac{\pi}{4}}2D$, fourth in $R{\frac{\pi}{2}}2D$, and last in $E{\frac{\pi}{2}}2D$. In all simulations $t_{sub}$ is close to $t_p$, i.e. the global maximum in the kinetic energy of the subharmonic approximately conincides with the initial crossing of pairing KH billows. \citet{Ho82} obtain a qualitatively similar result in the laboratory.

\section{Three-dimensionalization and mixing}\label{sec:3D}
\subsection{Three-dimensionalization}

The growth of three-dimensional instabilities inhibits pairing. In simulation $R\frac{\pi}{2}3D$, where pairing is delayed, three-dimensional instabilities break down the two individual KH billows before pairing can occur, see figure \ref{fig:vort_snap}, $t=146$. In simulation $R03D$ pairing occurs before the growth of three-dimensional instabilities and by $t=146$ the two vortices have merged into the subharmonic billow, see figure \ref{fig:vort_snap}. 

\begin{figure}
	\centerline{\includegraphics[width = \textwidth]{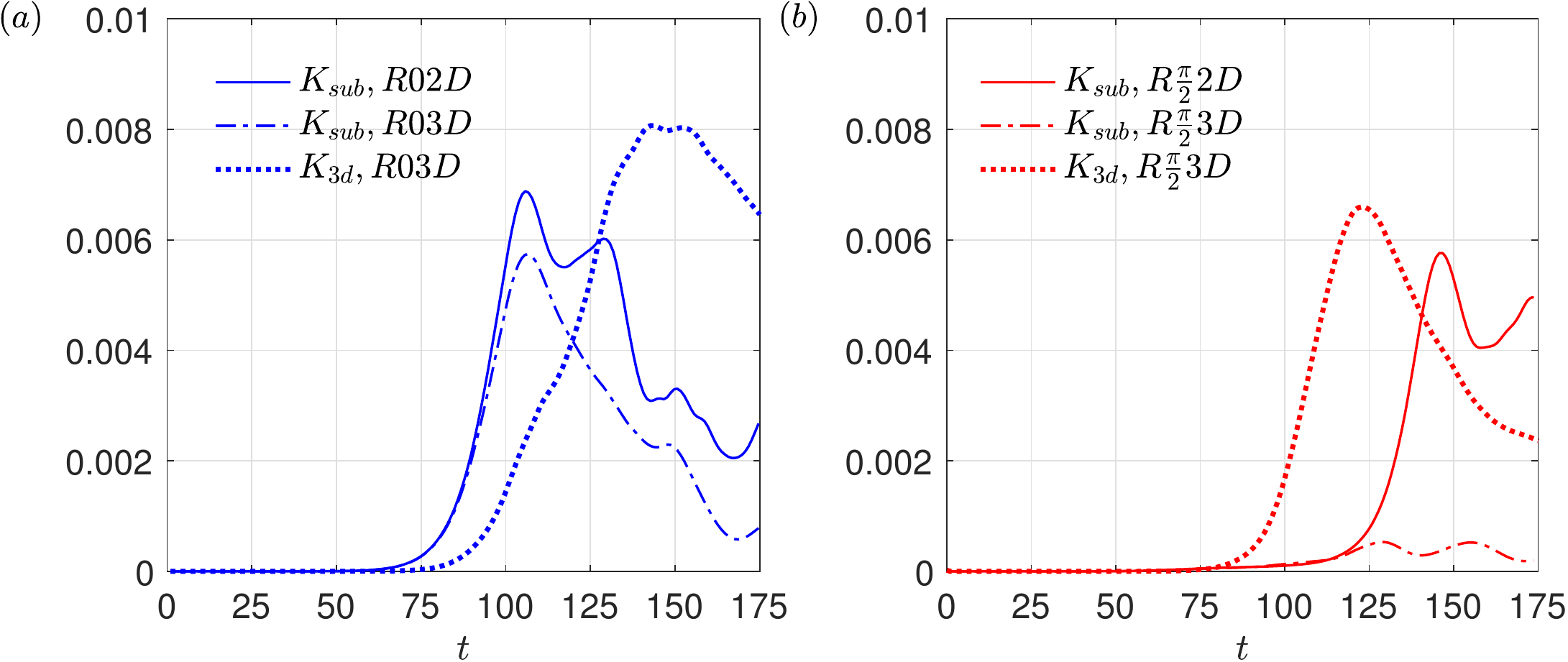}}
	\caption{Kinetic energy of the subharmonic mode $K_{sub}$ and three-dimensional kinetic energy $K_{3d}$ in two-dimensional and three-dimensional simulations: ($a$) $\theta_{sub}^M \approx 0$, ($b$) $\theta_{sub}^M \approx -\frac{\pi}{2}$.}
	\label{fig:ke3d}
\end{figure}

To quantify the effects of three-dimensional motions on pairing, we compare the kinetic energy of the subharmonic component in two- and three-dimensional random perturbation simulations to the kinetic energy of three-dimensional motions (figure \ref{fig:ke3d}) for most and least favourable phase conditions for vortex pairing. In the optimal phase simulation, i.e. $\theta_{sub}^M \approx 0$, the peak of the kinetic energy of the subharmonic mode, $K_{sub}$, is reduced slightly in the three-dimensional simulation, while the saturation time of the subharmonic mode is almost identical in the $2D$ and $3D$ simulations (see $R02D$ and $R03D$ in figure \ref{fig:ke3d} ($a$) and table \ref{tab:times}). The peak in $K_{3D}$ occurs at $t = 143$, well after the peak in $K_{sub}$. These indicate that in cases with the phase at or near optimal the growth of three-dimensional motions has little effect on pairing.

For $\theta_{sub}^M \approx -\frac{\pi}{2}$, the peak of the kinetic energy of the subharmonic mode $K_{sub}$ is significantly lower in the three-dimensional simulation compared to the two-dimensional simulation (see figure \ref{fig:ke3d} ($b$)). In the $R\frac{\pi}{2}3D$ simulation, the peak in $K_{3D}$ occurs earlier, at $t =123$, and precedes the peak of $K_{sub}$ in both two- and three-dimensional simulations. During the extra time needed in $R\frac{\pi}{2}3D$ for the phase to shift from $-\frac{\pi}{2}$ to $\sim 0$, the three-dimensional motions grow and and destroy the two-dimensional coherent KH billows before vortex pairing occurs. Therefore, pairing is eliminated in simulation $R\frac{\pi}{2}3D$.  The peak in $K_{3D}$ in simulation $R\frac{\pi}{2}3D$ is smaller compared to that in simulation $R03D$. The vortex pairing in simulation $R03D$ effectively increases the Reynolds number and makes the flow more energetic. 

\subsection{Mixing}\label{sec:mixing}
 
\begin{figure}
	\centerline{\includegraphics[width = 0.9\textwidth]{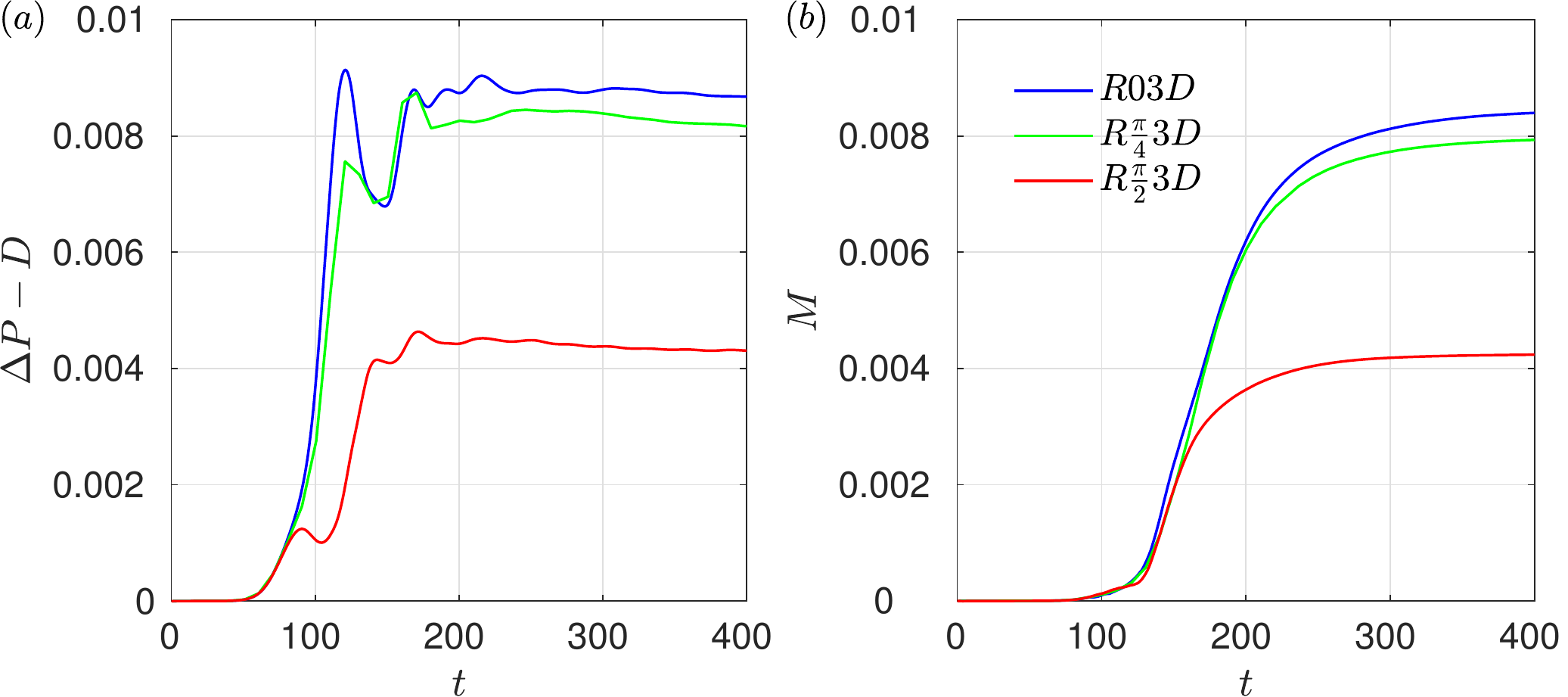}}
	\caption{(\textit{a}) The increase in total potential energy caused by fluid's motion $\Delta P - D$, (\textit{b}) the increase in mixing $M$.}
	\label{fig:potential_energy}
\end{figure}

Figure \ref{fig:potential_energy} ($a$) shows the increase in the total potential energy caused by the fluid's motion, $\Delta P - D$, with time. Time variation of $\Delta P -D$ is the same for the three simulations before  $t=80$. The first peak in $R03D$ and $R{\frac{\pi}{4}}3D$ represents vortex pairing  and the first peak in $R{\frac{\pi}{2}}3D$ represents saturation of KH. The peak due to vortex pairing does not exist for $R{\frac{\pi}{2}3}D$ as pairing never occurs. Contrary to the results of \citet{Mashayek13} where vortex pairing occurred during the turbulent stage, the peak of $\Delta P- D$ due to vortex pairing occurs during the pre-turbulent stage due to our relatively low Reynolds number.  Overall, the increase in total potential energy in cases with vortex pairing is much higher than the case without vortex pairing because vortex pairing efficiently increases the vertical scale of fluid's motion and stirs the fluid. The increase in total potential energy in $R{\frac{\pi}{4}}3D$ is also slightly lower than in $R03D$.

Figure \ref{fig:potential_energy} ($b$) shows the amount of mixing $M$. For all simulations, the amount of mixing is negligible before saturation of KH instabilities because the flow is still well-organized and mixing is mainly caused by laminar molecular diffusion. This is consistent with findings of \citet{Mashayek13}, \citet{Rahmani16} and  \citet{Salehipour15} which have shown that  mixing is negligible before the KH billow reaches its maximum amplitude.
After about $t=130$, the amount of mixing significantly increases as small-scale motions reach sufficient amplitude and mixing occurs through turbulent diffusion. As turbulence subsides, the amount of mixing gradually approaches a constant. 
 The final amount of mixing in simulation $R03D$ with vortex pairing is significantly higher than simulation $R\frac{\pi}{2}3D$ without pairing. The amount of mixing $M$ at $t=400$ in simulation $R03D$ 
is double that of simulation $R\frac{\pi}{2}3D$. 
 Mixing in simulation $R\frac{\pi}{4}3D$ is only slightly lower than that in simulation $R03D$. 
 
\begin{figure}
	\centerline{\includegraphics[width = 0.9\textwidth]{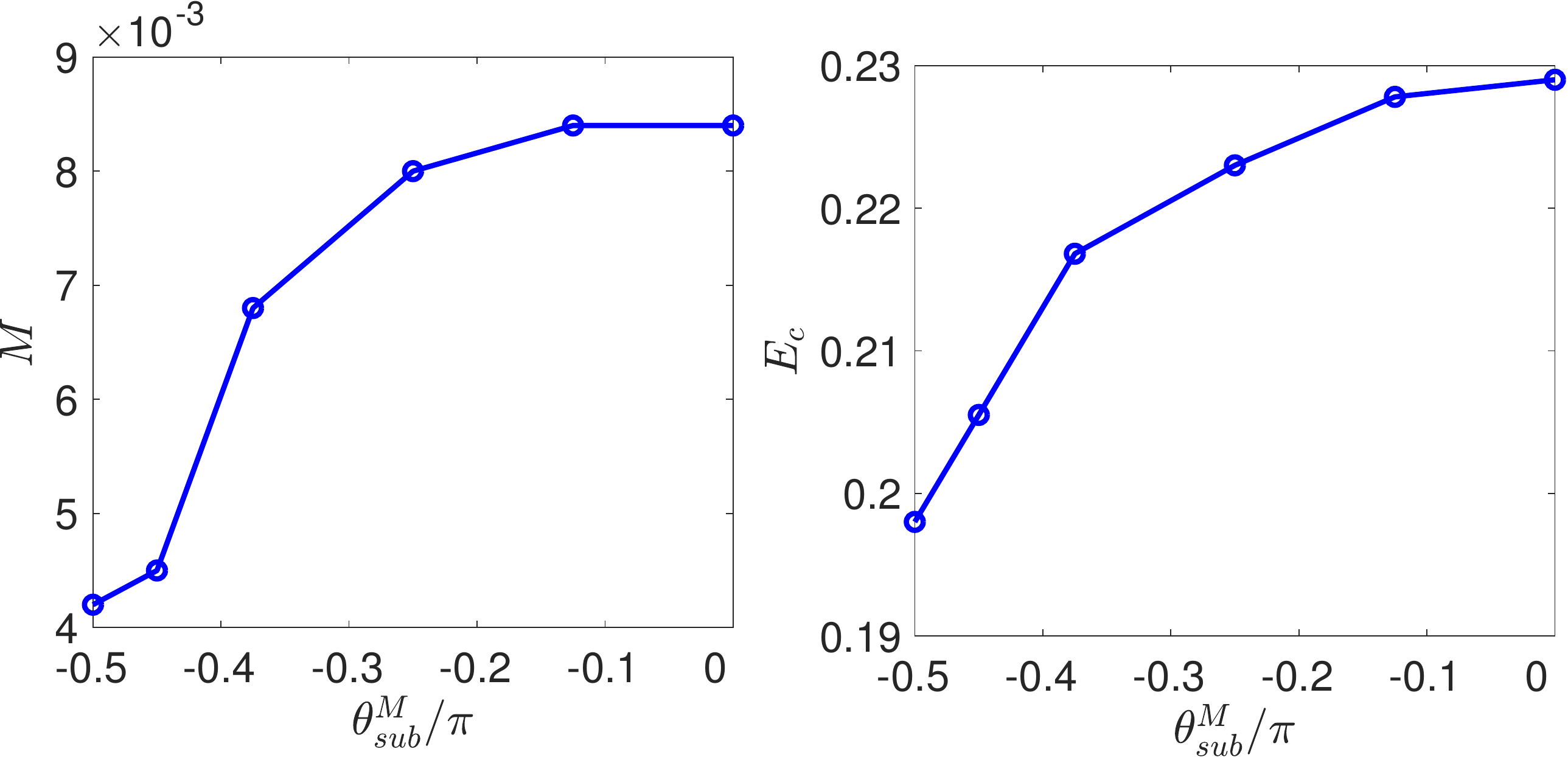}}
	\caption{The dependence of (a) the final amount of mixing $M$ and (b) the cumulative mixing efficiency $E_{c}$ on the phase difference between the primary KH and the subharmonic component, $\theta_{sub}^M$. }
	\label{fig:efficiency_mixing}
\end{figure}

We examine the dependence of the final amount of mixing $M$ and the cumulative mixing efficiency $E_c$, defined in equation (\ref{eqn:Ec}) as a measure of mixing efficiency during the active turbulence stage, on $\theta_{sub}^M$ in figure \ref{fig:efficiency_mixing}. As the phase of the subharmonic mode relative to KH decreases from $0$ to $-\dfrac{\pi}{2}$, mixing drops monotonically to less than half of its maximum value at $\theta_{sub}^M=0$. However, this effect is less pronounced when the phase difference is close to optimal and mixing starts to sharply drop for $\theta_{sub}^{M} \leq -0.375 \pi$. This is consistent with the laboratory experiments of  \cite{husain1995experiments} and \cite{hajj1993fundamental} where they observed the vortex pairing was suppressed over a range of phases close to the non-optimal phase. The cumulative mixing efficiency drops monotonically from 0.229 at $\theta_{sub}^M=0$ to 0.198 at $\theta_{sub}^M=-\dfrac{\pi}{2}$, marking a $14\%$ drop. Therefore, the effect of phase on cumulative mixing efficiency is less pronounced  compared to its effect on the amount of mixing. The range of $E_c$ obtained here is close to the cumulative mixing efficiency of 0.2 commonly computed in other numerical studies \citep{Smyth01,Peltier03,Rahmani14,Salehipour16,Rahmani16}. However, these studies have revealed some sensitivity of the mixing efficiency to the Reynolds number, Prandtl number and the bulk Richardson number that can make direct comparisons to our results less straightforward. 


\section{Discussion}\label{sec:discussion}

\begin{figure}
	\centerline{\includegraphics[width = 0.8\textwidth]{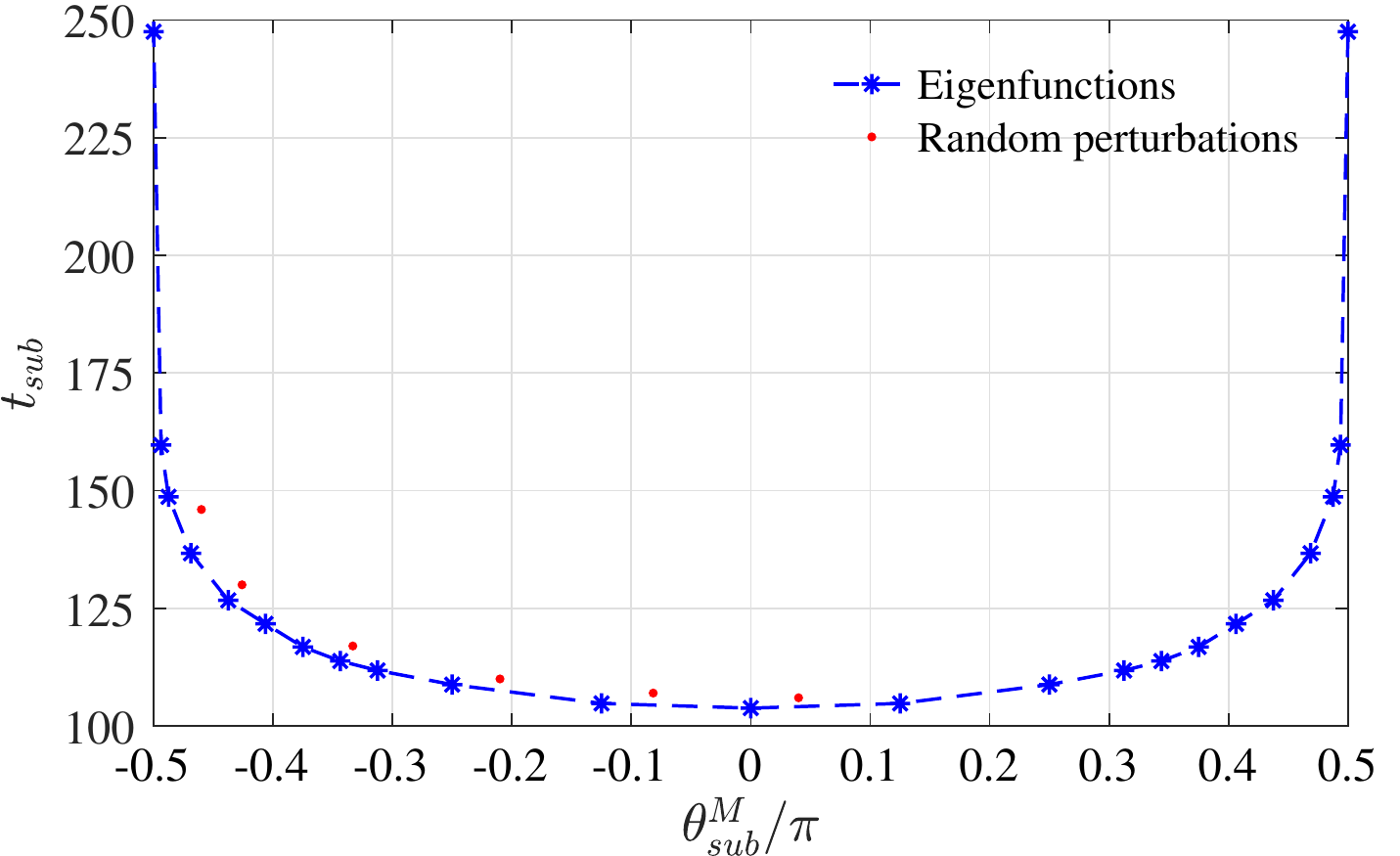}}
	\caption{Saturation time of the subharmonic mode, $t_{sub}$, as a function of the phase $\theta_{sub}^M$ for two-dimensional simulations perturbed by eigenfunctions and random perturbations. Note $t_{sub}$ is approximately equal to the time of pairing (see table \ref{tab:times}).} 
	\label{fig:tsub}
\end{figure}

We find that vortex pairing is sensitive to initial conditions when the phase of the subharmonic mode is close to $\pm \frac{\pi}{2}$. For simplicity in this discussion, we use $t_{sub}$ to characterize the time of pairing. Provided pairing occurs, this is generally accurate (i.e., $t_{sub}\sim t_p$). In general, $t_{sub}$ is a function of all modes in the initial conditions, not only the subharmonic mode. 

We consider the sensitivity of $t_{sub}$ to the phase of the subharmonic mode by running  two-dimensional simulations perturbed by KH and the subharmonic mode eigenfunctions.  Since the initial velocity and density fields of phase $\theta_{sub}^M$ are the negatives of the reflection of velocity and density fields of phase $\theta_{sub}^M$ about the centre of the domain and the governing equations conserve this symmetry,  $t_{sub}$ is  an even function of $\theta_{sub}^M$. Therefore, we only consider phases from $-\frac{\pi}{2}$ to 0 for random perturbations. We calculate $t_{sub}$ for 12 discrete phases between $-\frac{\pi}{2}$ and 0 and plot the results and the symmetric reflection between 0 and $\frac{\pi}{2}$ in figure \ref{fig:tsub}. The figure shows that the time of pairing is not sensitive to the phase when the phase is close to 0. However, the time of pairing increases significantly near $\theta_{sub}^M=\pm \frac{\pi}{2}$, which indicates that $t_{sub}$ is sensitive to the phase when the phase is close to $\pm \frac{\pi}{2}$. 
 Additional simulations (not shown) indicate that slight deviation from the eigenfunction of the subharmonic mode in initial conditions can also change the time of pairing. Hence, the delay of vortex pairing is sensitive to the functional form of the subharmonic component, phase, and other modes in two-dimensional simulations when the phase is close to $\pm \frac{\pi}{2}$. Detailed investigations of these effects are the subject of future studies. 

 Figure \ref{fig:tsub} also shows the time of pairing $t_{sub}$ for two-dimensional random perturbation simulations. The results show the same trend as the eigenfunction simulations, i.e., $t_{sub}$ increases with the phase. The initial non-modal growth and existence of other modes in the initial conditions cause the slight difference between random perturbation  and eigenfunction results.  However, the significant increase of $t_{sub}$ near $\pm \frac{\pi}{2}$ can only occur in pure eigenfunction simulations since any deviation from the pure eigenfunctions in initial conditions will project on the pairing mode with phase 0 and reduce the time of pairing compared to pure eigenfunctions with phase $\pm \frac{\pi}{2}$.

\section{Conclusions} \label{sec:conclusion}
We have investigated the effect of phase of subharmonic mode on vortex pairing and mixing using two-dimensional and three-dimensional DNSs.  In two-dimensional simulations, we use a ratio to measure the extent to which that the subharmonic component deviates from the eigenfunction to the TG equation with the same wavenumber. That the ratio quickly increases to 1 from a small number shows that the non-modal subharmonic component quickly evolves to the eigenfunction. We also track the Lagrangian trajectories of two fluid particles located at the centres of the KH vortices and their trajectories are shown to represent the vortex centres before small scale motions prevail. Similar to \citet{Ho82}, when kinetic energy of the subharmonic mode reaches its maximum, one KH vortex is almost on top of the other, i.e. $t_{sub}$ coincides with $t_p$. 

As \citet{Klaassen89} and \citet{Smyth93} have shown, if the subharmonic mode is out of phase, it adjusts its phase and pairing is delayed. We have found that if the initial phase of the subharmonic mode is not close to $\pm \frac{\pi}{2}$, pairing is only slightly delayed and the flow perturbed by eigenfunctions behaves similarly to the flow perturbed by random perturbations.  Before the KH instability reaches its first maximum in kinetic energy, i.e. before $t_{kh}$, the growth rate of the subharmonic component is almost the same for the eigenfunction simulation and random perturbation simulation if the phase of the subharmonic component is the same. Moreover, the growth rate in the case where the phase is about $\frac{\pi}{4}$ is close to the case where the phase is 0. After $t_{kh}$, if the phase is not close to $\pm \frac{\pi}{2}$, the growth rate is still not sensitive to the initial perturbations. If the phase is close to $\pm \frac{\pi}{2}$, the growth rate in the random perturbation differs significantly from that in the eigenfunction simulation and the subharmonic component reaches its maximum earlier in the random perturbation simulation. To investigate the sensitivity of time of pairing to the phase, we ran simulations perturbed by the KH and subharmonic eigenfunctions with different phases. It is shown that $t_{sub}$ increases with the phase of the subharmonic mode and it increases significantly at $\theta_{sub}^M = \pm \frac{\pi}{2}$. Time of pairing and $t_{sub}$ show the same trend in the flows perturbed by eigenfunctions as in the flows perturbed by random perturbations.  


In three-dimensional simulations, vortex pairing is always suppressed by three-dimensional motions and the suppression is greater when the phase difference is larger. Thus the maximum two-dimensional kinetic energy decreases as the phase increases. Three-dimensional motions can grow to sufficient amplitude and eliminate pairing when the phase difference is sufficiently large. Weaker pairing leads to less mixing inasmuch as mixing for the phase of the subharmonic mode of $ -\frac{\pi}{2}$ mixing drops more than two times compared to when the phase is $0$. The mixing efficiency however diminishes only slightly when the phase changes from $0$ to $ -\frac{\pi}{2}$ and its value remains close to 0.2, commonly found in previous studies. Mixing sharply decreases as the phase approaches $- \frac{\pi}{2}$, similar to the sharp increase in the time of pairing close to $\pm \frac{\pi}{2}$. These results are consistent with the laboratory observations of \cite{hajj1993fundamental}, and \cite{husain1995experiments} for the suppression of the subharmonic mode close to an unfavourable phase. \\

We are thankful to Dr. Kraig Winters and Dr. Bill Smyth for providing the DNS code and
Westgrid for computational resources.
\bibliographystyle{jfm}
\bibliography{rapids}

\newpage

\end{document}